\documentclass{llncs}

\usepackage{url}
\usepackage{a4wide}

\usepackage{xspace}
\usepackage{mathtools}

\newcommand\refthm[1]{Theorem~\ref{#1}\xspace} 
\newcommand\reflma[1]{Lemma~\ref{#1}\xspace} 
\newcommand\refsec[1]{Section~\ref{#1}\xspace} 
\newcommand\etal{~\textit{et al}.\xspace}

\newcommand\sideheading[1]{\medskip\noindent{\textbf{#1}}\xspace}

\newcommand\recaplma[2]{
        \longshort{
            \renewcommand\thelemma{\ref{#1}}
        }{
            \renewcommand\thetheorem{\ref{#1}}
        }
        \begin{lemma}
            {#2}
        \end{lemma}
        \addtocounter{theorem}{-1}
}

\newcommand\brac[1]{\left({#1}\right)} \newcommand\tuple[1]{\brac{#1}} 
\newcommand\set[1]{\left\{\ {#1}\ \right\}}
\newcommand\setcomp[2]{\left\{\ {#1}\ \left|\ {#2}\ \right.\right\}} 
\newcommand\order[1]{{\mathcal{O}\mathord{\brac{{#1}}}}}
\newcommand\asize[1]{{\left|{#1}\right|}}

\newcommand\PDS{PDS\xspace}
\newcommand\PDSs{PDSs\xspace}
\newcommand\NFA{NFA\xspace}
\newcommand\CFG{CFG\xspace}
\newcommand\CFGs{CFGs\xspace}

\newcommand\naPDS{naPDS\xspace}
\newcommand\naPDSs{naPDSs\xspace}
\newcommand\NPDS{NPDS\xspace}
\newcommand\NPDSs{NPDSs\xspace}

\newcommand\pds{\mathcal{P}}
\newcommand\npds{\mathcal{N}}
\newcommand\galphabet{\mathcal{G}} 
\newcommand\controls{\mathcal{Q}}
\newcommand\salphabet{\Sigma}
\newcommand\oalphabet{\Gamma}
\newcommand\pdsrules{\Delta}
\newcommand\gread[1]{{r\mathord{\brac{{#1}}}}}
\newcommand\gwrite[1]{{w\mathord{\brac{{#1}}}}} 
\newcommand\config[2]{\tuple{{#1}, {#2}}}
\newcommand\pdsrulebasic[5]{\tuple{{#1}, {#2}} \xhookrightarrow{{#3}} \tuple{{#4}, {#5}}}
\newcommand\pdsrule[4]{\pdsrulebasic{#1}{#2}{}{#3}{#4}}
\newcommand\pdsruler[5]{\pdsrulebasic{#1}{#2}{\gread{{#3}}}{#4}{#5}}
\newcommand\pdsrulew[5]{\pdsrulebasic{#1}{#2}{\gwrite{{#3}}}{#4}{#5}} 
\newcommand\pdstranbasic[1]{\xrightarrow{{#1}}}
\newcommand\pdstran{\pdstranbasic{}}
\newcommand\pdstranr[1]{\pdstranbasic{\gread{{#1}}}}
\newcommand\pdstranw[1]{\pdstranbasic{\gwrite{{#1}}}}
\newcommand\localconfig[2]{\tuple{{#1}, {#2}}}
\newcommand\npdstran{\pdstran}
\newcommand\master{\mathcal{U}}
\newcommand\slave{\mathcal{C}}
\newcommand\run{\pi}
\newcommand\sbot{\perp}
\newcommand\killg{\#}
\newcommand\readchars{R}

\newcommand\langof[1]{\mathcal{L}\mathord{\brac{{#1}}}}
\newcommand\lang{\mathcal{L}} 

\newcommand\states{\mathcal{Q}}
\newcommand\transitions{\Delta} 
\newcommand\finals{\mathcal{F}}
\newcommand\nfa{\mathcal{A}}
\newcommand\transition[3]{{#1} \xrightarrow{{#2}} {#3}}
\newcommand\runtransition[1]{\xrightarrow{{#1}}}

\newcommand\parampds{\pds_{sys}}

\newcommand\cfg{G} 
\newcommand\hasspine[2]{{\delta\mathord{\brac{{#1},{#2}}}}}
\newcommand\spinetype[1]{{\theta\mathord{\brac{{#1}}}}}
\newcommand\wordtype[2]{{\tau\mathord{\brac{{#1},{#2}}}}}
\newcommand\numchar[2]{{\#_{{#1}}\mathord{\brac{{#2}}}}}
\newcommand\dtree{T}
\newcommand\marked[1]{\overline{{#1}}} 
\newcommand\lalphabet{\Sigma} 
\newcommand\prodarr{\rightarrow} 
\newcommand\dpath{\rho} 
\newcommand\dlab[1]{{\ell\mathord{\brac{{#1}}}}}
\newcommand\dmarklab[1]{{\marked{\ell}\mathord{\brac{{#1}}}}}
\newcommand\markedtree[2]{{\marked{{#2}}\mathord{\brac{{#1}}}}} 
\newcommand\spine[1]{{Spine\mathord{\brac{{#1}}}}}
\newcommand\typedg[2]{{{#1}_{#2}}}
\newcommand\typenfa[1]{{\nfa_{{#1}}}}

\newcommand\longshort[2]{#1}
\newcommand\forjade[2]{#1}

\begin{document}
   
    \title{Parameterised Pushdown Systems with Non-Atomic Writes}
    \author{M. Hague}
    
    \institute{
        Oxford University, Department of Computer Science \\
        and \\
        Laboratoire d'Informatique Gaspard-Monge, Universit\'e Paris-Est
    }
    \date{}

    \maketitle
    
    \begin{abstract}

We consider the master/slave parameterised reachability problem for networks of
pushdown systems, where communication is via a global store using only
non-atomic reads and writes.  We show that the control-state reachability
problem is decidable.  As part of the result, we provide a constructive
extension of a theorem by Ehrenfeucht and Rozenberg to produce an \NFA
equivalent to certain kinds of \CFG.  Finally, we show that the
non-parameterised version is undecidable.

\longshort{     
    \medskip

    Note, this is the long version of work appearing in FSTTCS 2011.
}{}
    
    \end{abstract}

\section{Introduction}

A parameterised reachability problem is one where the system is defined in terms
of a given input, usually a number $n$.  We then ask whether there is some $n$
such that the resulting system can reach a given state.  An early result shows
that this problem is undecidable, even when the system defined for each $n$ is a
finite state machine: one simply has to define the $n$th system to simulate a
Turing machine up to $n$ steps~\cite{AK86}.  Thus, the Turing machine terminates
iff there is some $n$ such that the $n$th system reaches a halting state.

Such a result, however, is somewhat pathological.  More natural parameterised
problems concentrate on the replication of components.  For instance, we may
have a leadership election algorithm amongst several nodes.  For this algorithm
we would want to know, for example, whether there is some $n$ such that, when
$n$ nodes are present, the routine fails to elect a leader.  This problem walks
the line between decidability and undecidability, even with finite-state
components: in a ring network, when nodes can communicate to their left and
right neighbours directly, Suzuki proves undecidability~\cite{S88};  but, in
less disciplined topologies, the problem becomes decidable~\cite{GS92}.

In particular, the above decidability result considers the following problem:
given a master process $\master$ and slave $\slave$, can the master in parallel
with $n$ slaves reach a given state.  Communication in this system is by
anonymous pairwise synchronisation (that is, a receive request can be satisfied
by \emph{any} thread providing the matching send, rather than a uniquely
identified neighbour).  This problem reduces to Petri-nets, which can, for each
state of $\slave$, keep a count of the number of threads in that state.  When
communication is via a finite-state global store, which all threads can read
from and write to (atomically), it is easy to see that decidability can be
obtained by the same techniques.

These results concern finite-state machines.  This is ideal for hardware or
simple protocols.  When the components are more sophisticated (such as threads
created by a web-server), a more natural and expressive (infinite-state) program
model --- allowing one to accurately simulate the control flow of first-order
recursive programs~\cite{JM77} --- is given by pushdown systems (\PDSs).  Such
systems have proved popular in the sequential setting
(e.g.~\cite{BEM97,EKS01,S02,RSJD05}), with several successful
implementations~\cite{BR00,BR02,S02}.  Unfortunately, when two \PDSs
can communicate, reachability quickly becomes undecidable~\cite{R00}.

In recent years, many researchers have tackled this problem, proposing many
different approximations, and restrictions on topology and communication
behaviour (e.g.~\cite{M98,BESS05,BET03,BMT05,SV06,QR05,HLMS10}).  A pleasantly
surprising (and simple) result in this direction was provided by
Kahlon~\cite{K08}: the parameterised reachability problem for systems composed
of $n$ slaves $\slave$ communicating by anonymous synchronisation is decidable.
This result relies heavily on the inability of the system to restrict the number
of active processes, or who they communicate with.  Indeed, in the presence of a
master process $\master$, or communication via a global store, undecidability is
easily obtained.  

In this work we study the problem of adding the master process and global store.
To regain decidability, we only allow \emph{non-atomic} accesses to the shared
memory.  We then show --- by extending a little-cited theorem of Ehrenfeucht and
Rozenberg~\cite{ER85} --- that we can replace the occurrences of $\slave$ with
regular automata\footnote{A reviewer points out that the upward-closure of a
context free language has been proved regular by Atig\etal~\cite{ABT08} with the
same complexity, which is sufficient for our purposes.  However, a constructive
version of Ehrenfeucht and Rozenberg is a stronger result, and hence remains a
contribution.}.  This requires the introduction of different techniques than
those classically used.  Finally, a product construction gives us our result.
In addition, we show that, when $n$ is fixed, the problem remains undecidable,
for all $n$.  For clarity, we present the single-variable case here.  In the
appendix we show that the techniques extend easily to the case of $k$ shared
variables.


After discussing further related work, we begin in \refsec{sec:preliminaries}
with the preliminaries.  In \refsec{sec:model} we define the systems that we
study.  Our main result is given is \refsec{sec:result} and the accompanying
undecidability proof appears in \refsec{sec:undecidability}.  In
\refsec{sec:cfg2reg} we show how to obtain a constructive version of Ehrenfeucht
and Rozenberg's theorem.  Finally, we conclude in \refsec{sec:conclusion}.
\longshort{}{A version of this paper complete with appendix is available from
\url{http://www.cs.ox.ac.uk/matthew.hague/fsttcs11}.}

\sideheading{Related Work} 
Many techniques attack parameterisation (e.g.\ network invariants and symmetry).
Due to limited space, we only discuss \PDSs here.  In addition to results on
parameterised \PDSs, Kahlon shows decidability of concurrent \PDSs communicating
via nested-locks~\cite{KIG05}.  In contrast, we cannot use locks to guarantee
atomicity here.

A closely related model was studied by Bouajjani\etal in 2005.  As we
do, they allow \PDSs to communicate via a global store.  They do not consider
parameterised problems directly, but they do allow the dynamic
creation of threads.  By dynamically creating an arbitrary number of threads at
the start of the execution, the parameterised problem can be simulated.
Similarly, parameterisation can simulate thread creation by activating hitherto
dormant threads.  However, since Bouajjani\etal allow atomic read/write actions
to occur, the problem they consider is undecidable; hence, they consider
\emph{context-bounded reachability}.

Context-bounded reachability is a popular technique based on the observation
that many bugs can be identified within a small number of context
switches~\cite{Q08}.  This idea has been extended to \emph{phase-bounded}
systems where only one stack may be decreasing in any one
phase~\cite{ABH08,S10}.  Finally, in another extension of context-bounded
model-checking, Ganty\etal consider \emph{bounded under-approximations} where
runs are restricted by intersecting with a word of the form $a_1^\ast\ldots
a_n^\ast$~\cite{GMM10}.  In contrast to this work, these techniques are only
accurate up to a given bound.  That is, they are sound, but not complete.
Recently, La Torre\etal gave a sound algorithm for parameterised \PDSs together
with a technique that may detect completeness in the absence of
recursion~\cite{lTMP10}.

Several models have been defined for which model-checking can be sound and
complete.  For example, Bouajjani\etal also consider acyclic
topologies~\cite{ABT08,BMT05}.  As well as restricting the network structure,
Sen and Viswanathan~\cite{SV06}, La Torre\etal~\cite{lTMP08} and later
Heu{\ss}ner\etal~\cite{HLMS10}, show how to obtain decidability by only allowing
communications to occur when the stack satisfies certain conditions.

One of the key properties that allow parameterized problems to become decidable
is that once a copy of the duplicated process has reached a given state, then
any number of additional copies may also be in that state.  In effect, this
means that any previously seen state may be returned to at any time.  This
property has also been used by Delzanno\etal to analyse recursive ping-pong
protocols~\cite{DES06} using \emph{Monotonic Set-extended Prefix Rewriting}.
However, unlike our setting, these systems do not have a master process.

Finally, recent work by Abdulla\etal considers parameterised problems with
non-atomic global conditions~\cite{AHDR08}.  That is, global transitions may
occur when the process satisfy a global condition that is not evaluated
atomically.  However, the processes they consider are finite-state in general.
Although a procedure is proposed when unbounded integers are allowed, this is
not guaranteed to terminate.

\section{Preliminaries} \label{sec:preliminaries}

We recall the definitions of finite automata and pushdown systems and their
language counter-parts.  We also state a required result by Ehrenfeucht and
Rozenberg.

\begin{definition}[Non-Deterministic Finite Word Automata]
    We define a \emph{non-deterministic finite word automaton} (\NFA) $\nfa$ as
    a tuple $\tuple{\states, \oalphabet, \transitions, q_0, \finals}$ where
    $\states$ is a finite set of states, $\oalphabet$ is a finite alphabet, $q_0
    \in \states$ is an initial state, $\finals \subseteq \states$ is a set of
    final states, and $\transitions \subseteq \states \times \oalphabet \times
    \states$ is a finite set of transitions.  
\end{definition} 
We will denote a transition $\tuple{q, \gamma, q'}$ using the notation
$\transition{q}{\gamma}{q'}$.  We call a sequence $q_1 \runtransition{\gamma_1}
q_2 \runtransition{\gamma_2} \cdots \runtransition{\gamma_{z-1}} q_z$ a
\emph{run} of $\nfa$.  It is an accepting run if $q_1 = q_0$ and $q_z \in
\finals$.  The language $\langof{\nfa}$ of an \NFA is the set of all words
labelling an accepting run.  Such a language is \emph{regular}.

\begin{definition}[Pushdown Systems] 
    A \emph{pushdown system} (\PDS) $\pds$ is \longshort{}{defined as} a tuple
    $\tuple{\controls, \salphabet, \oalphabet, \pdsrules, q_0, \finals}$ where
    $\controls$ is a finite set of control states, $\salphabet$ is a finite
    stack alphabet with a special bottom-of-stack symbol $\sbot$, $\oalphabet$
    is a finite output alphabet, $q_0 \in \controls$ is an initial state,
    $\finals \subseteq \controls$ is a set of final states, and $\pdsrules
    \subseteq \brac{\controls \times \salphabet} \times \oalphabet \times
    \brac{\controls \times \salphabet^\ast}$ is a finite set of transition
    rules.  
\end{definition}
We will denote a transition rule $((q, a), \gamma, (q', w'))$ using the notation
$\pdsrulebasic{q}{a}{\gamma}{q'}{w'}$.  The bottom-of-stack symbol is neither
pushed nor popped.  That is, for each rule $\pdsrulebasic{q}{a}{\gamma}{q'}{w'}
\in \pdsrules$ we have, when $a \neq \sbot$, $w$ does not contain $\sbot$, and,
$a = \sbot$ iff $w' = w\sbot$ and $w$ does not contain $\sbot$.  A configuration
of $\pds$ is a tuple $\config{q}{w}$, where $q \in \controls$ is the current
control state and $w \in \salphabet^\ast$ is the current stack contents.  There
exists a transition $\config{q}{aw} \pdstranbasic{\gamma} \config{q'}{w'w}$ of
$\pds$ whenever $\pdsrulebasic{q}{a}{\gamma}{q'}{w'} \in \pdsrules$.  We call a
sequence $c_0 \pdstranbasic{\gamma_1} c_1 \pdstranbasic{\gamma_2} \cdots
\pdstranbasic{\gamma_z} c_z$ a \emph{run} of $\pds$.  It is an accepting run if
$c_0 = \config{q_0}{\sbot}$ and $c_z = \config{q}{w}$ with $q \in \finals$.  The
language $\langof{\pds}$ of a pushdown system is the set of all words labelling
an accepting run.  Such a language is \emph{context-free}.  Note, in some cases,
we omit the output alphabet $\oalphabet$.  In this case, the only character is
the empty character $\varepsilon$, with which all transitions are labelled.  In
general, we will omit the empty character $\varepsilon$ when it labels a
transition.

We use a theorem of Ehrenfeucht and Rozenberg~\cite{ER85}.  With respect to a
context-free language $\lang$, a \emph{strong iterative pair} is a tuple
$\tuple{x, y, z, u, t}$ of words such that for all $i \geq 0$ we have $x y^i z
u^i t \in \lang$, where $y$ and $u$ are non-empty words.  A strong iterative
pair is \emph{very degenerate} if, for all $i, j \geq 0$ we have that $x y^i z
u^j t \in \lang$.
\begin{theorem}[\cite{ER85}] \label{thm:very-degenerate}
    For a given context-free language $\lang$, if all strong iterative pairs are
    very degenerate, then $\lang$ is regular.
\end{theorem}
However, Ehrenfeucht and Rozenberg do not present a constructive algorithm for
obtaining a regular automaton accepting the same language as an appropriate
context-free language.  Hence, we provide such an algorithm in
\refsec{sec:cfg2reg}.

\section{Non-Atomic Pushdown Systems} \label{sec:model}

Given an alphabet $\galphabet$, let $\gread{\galphabet} = \setcomp{\gread{g}}{g
\in \galphabet}$ and $\gwrite{\galphabet} = \setcomp{\gwrite{g}}{g \in
\galphabet}$.  These alphabets represent read and write actions respectively of
the value $g$.

\begin{definition}[Non-atomic Pushdown Systems]
    Over a finite alphabet $\galphabet$, a \emph{non-atomic pushdown system}
    (\naPDS) is a tuple $\pds = \tuple{\controls, \salphabet, \pdsrules, q_0,
    \galphabet}$ where $\controls$ is a finite set of control-states,
    $\salphabet$ is a finite stack alphabet with a bottom-of-stack symbol
    $\sbot$, $q_0 \in \controls$ is a designated initial control state and
    $\pdsrules \subseteq \brac{\controls \times \salphabet} \times
    \brac{\gread{\galphabet} \cup \gwrite{\galphabet} \cup \set{\varepsilon}}
    \times \brac{\controls \times \salphabet^\ast}$.
\end{definition}

That is, a non-atomic pushdown system is a \PDS where the output alphabet is
used to signal the interaction with a global store, and there are no final
states: we are interested in the behaviour of the system, rather than the
language it defines.

\begin{definition}[Networks of \naPDSs]
    A network of $n$ \emph{non-atomic pushdown systems} (\NPDS) is a tuple
    $\npds = \tuple{\pds_1, \ldots, \pds_n, \galphabet, g_0}$ where, for all $1
    \leq i \leq n$, $\pds_i = \tuple{\controls_i, \salphabet_i, \pdsrules_i,
    q^i_0, \galphabet}$ is a $\NPDS$ over $\galphabet$ and $g_0 \in \galphabet$
    is the initial value of the global store.
\end{definition}

A configuration of an \NPDS is a tuple $\config{q_1, w_1, \ldots, q_n, w_n}{g}$
where $g \in \galphabet$ and for each $i$, $q_i \in \controls_i$ and $w_i \in
\Sigma^\ast_i$.  There is a transition $\config{q_1, w_1, \ldots, q_n, w_n}{g}
\npdstran \config{q'_1, w'_1, \ldots, q'_n, w'_n}{g'}$ whenever, for some $1
\leq i \leq n$ and all $1 \leq j \leq n$ with $i \neq j$, we have $q'_j = q_j$,
$w'_j = w_j$, and
\begin{itemize}
    \item $\localconfig{q_i}{w_i} \pdstran \localconfig{q'_i}{w'_i}$ is a
          transition of $\pds_i$ and $g' = g$; or

    \item $\localconfig{q_i}{w_i} \pdstranr{g} \localconfig{q'_i}{w'_i}$ is a
          transition of $\pds_i$ and $g' = g$; or

    \item $\localconfig{q_i}{w_i} \pdstranw{g'} \localconfig{q'_i}{w'_i}$ is a
          transition of $\pds_i$.
\end{itemize}
A path $\run$ of $\npds$ is a sequence of configurations $c_1 c_2 \ldots c_m$
such that, for all $1 \leq i < m$, $c_i \npdstran c_{i+1}$.  A run of $\npds$ is
a path such that $c_1 = \config{q^1_0, \sbot, \ldots, q^n_0, \sbot}{g_0}$.

\section{The Parameterised Reachability Problem} \label{sec:result}

We define and prove decidability of the parameterised reachability problem for
\naPDSs.  We finish with a few remarks on the extension to multiple variables,
and on complexity issues.

\begin{definition}[Parameterised Reachability]
    For given \naPDSs $\master$ and $\slave$ over $\galphabet$, initial store
    value $g_0$ and control state $q$, the parameterised reachability problem
    asks whether there is some $n$ such that the \NPDS $\npds_n =
    \tuple{\master,\underbrace{\slave,\ldots,\slave}_{n},\galphabet, g_0}$ has a
    run to some configuration containing the control state $q$.    
\end{definition}

In this section, we aim prove the following theorem.
\begin{theorem}
    The parameterised reachability problem for \NPDSs is decidable.
\end{theorem}

Without loss of generality, we can assume $q$ is a control-state of $\master$ (a
$\slave$ process can write its control-state to the global store for $\master$
to read).  The idea is to build an automaton which describes for each $g \in
\galphabet$ the sequences $g_1\ldots g_m \in \galphabet^\ast$ that need to be
read by some $\slave$ process to be able to write $g$ to the global store.  We
argue using \refthm{thm:very-degenerate} that such \emph{read languages} are
regular (and construct regular automata using \reflma{lma:cfg2nfa}).  Broadly
this is because, between any two characters to be read, any number of characters
may appear in the store and then be overwritten before the process reads the
required character.  We then combine the resulting languages with $\master$ to
produce a context-free language that is empty iff the control-state $q$ is
reachable.

\subsection{Regular Read Languages} \label{sec:regread}

For each $g \in \galphabet$ we will define a \emph{read-language}
$\lang_\gwrite{g}$ which intuitively defines the language of read actions that
$\slave$ must perform before being able to write $g$ to the global store.  Since
$\slave$ may have to write other characters to the store before $g$, we use the
symbol $\killg$ as an abstraction for these writes.  The idea is that, for any
run of the parameterised system, we can construct another run where each copy of
$\slave$ is responsible for a single particular write to the global store, and
$\lang_\gwrite{g}$ describes what $\slave$ must do to be able to write $g$.  

To this end, given a non-atomic pushdown system $\pds$ we define for each $g \in
\galphabet$ the pushdown system $\pds_{\gwrite{g}}$ which is $\pds$ augmented
with a new unique control-state $f$, and a transition $\pdsrule{q}{a}{f}{a}$
whenever $\pds$ has a rule $\pdsrulew{q}{a}{g}{q'}{w}$.  Furthermore, replace
all $\pdsrulew{q}{a}{g'}{q'}{w}$ rules with $\pdsrulebasic{q}{a}{\killg}{q'}{w}$
where $\killg \notin \galphabet$.  These latter rules signify that the global
store contents have been changed, and that a new value must be written before
reading can continue.  This implicitly assumes that $\slave$ does not try to
read the last value it has written.  This can be justified since, whenever this
occurs, because we are dealing with the parameterised version of the problem, we
can simply add another copy of $\slave$ to produce the required write.

We interpret $f$ as the sole accepting control state of $\pds_{\gwrite{g}}$ and
thus $\langof{\pds_\gwrite{g}}$ is the language of reads (and writes) that must
occur for $g$ to be written.  We then allow any number of (ignored) read and
$\killg$ events\footnote{Extra $\killg$ events will not allow spurious runs, as
they only add extra behaviours that may cause the system to become stuck.  This
is because $\killg$ is never read by a process.} to occur.  That is, any word in
the read language contains a run of $\slave$ with any number of additional
actions that do not affect the reachability property interspersed.  Let
$\readchars = \setcomp{\gread{g'}}{g' \in \galphabet} \cup \set{\killg}$, we
define the read language $\lang_\gwrite{g} \subseteq \readchars^\ast$ for
$\gwrite{g}$ as
\[ 
    \lang_\gwrite{g} = \setcomp{\readchars^\ast\gamma_1\readchars^\ast\ldots
    \readchars^\ast\gamma_z\readchars^\ast}{\gamma_1\ldots\gamma_z \in
    \langof{\pds_\gwrite{g}}} \ .
\]
Note, in particular, that $\gamma_1\ldots\gamma_z \in \readchars^\ast$.

\begin{lemma}
    For all $g \in \galphabet$, $\lang_\gwrite{g}$ is regular and an \NFA $\nfa$
    accepting $\lang_\gwrite{g}$, of doubly-exponential size, can be constructed
    in doubly-exponential time.    
\end{lemma}
\begin{proof}
    Take any strong iterative pair $\tuple{x, y, z, t, u}$ of
    $\lang_\gwrite{g}$.  To satisfy the preconditions of
    \refthm{thm:very-degenerate}, we observe that $xzu \in \lang_\gwrite{g}$
    since we have a strong iterative pair.  Then, from the definition of
    $\lang_\gwrite{g}$ we know $x \readchars^\ast z \readchars^\ast u \subseteq
    \lang_\gwrite{g}$ and hence, for all $i, j$, $x y^i z t^j u \subseteq
    \lang_\gwrite{g}$ as required.  Thus $\lang_\gwrite{g}$ is regular.  The
    construction of $\nfa$ comes from \reflma{lma:cfg2nfa}.
\end{proof}

\subsection{Simulating the System}

We build a PDS that recognises a non-empty language iff the parameterised
reachability problem has a positive solution.  The intuition behind the
construction of $\parampds$ is that, if a collection of $\slave$ processes have
been able to use the output of $\master$ to produce a write of some $g$ to the
global store, then we may reproduce that group of processes to allow as many
writes $g$ to occur as needed.  Hence, in the construction below, once $q_i \in
\finals_i$ has been reached, $g_i$ can be written at any later time.  The
$\killg$ character is used to prevent sequences such as
$\gread{g}\gwrite{g'}\gread{g}$ occurring in read languages, where no process is
able to provide the required write $\gwrite{g}$ that must occur after
$\gwrite{g'}$.  Note that, if we did not use $\killg$ in the read languages,
such sequences could occur because the $\gwrite{g'}$ would effectively be
ignored.

The construction itself is a product construct between $\master$ and the regular
automata accepting the read languages of $\slave$.  The regular automata read
from the global variable, writing $\killg$ when a $\killg$ action should occur.
Essentially, they mimic the behaviour of an arbitrary number of $\slave$
processes in their interaction --- via the global store --- with $\master$ and
each other.  The value of the global store is held in the last component of the
product.

\begin{definition}[$\parampds$]
    Given an \naPDS $\master = \tuple{\controls_\master, \salphabet,
    \pdsrules_\master, q^\master_0, \galphabet}$ with initial store value $g_0$,
    a control-state $f \in \controls_\master$, and, for each $g \in \galphabet$,
    a regular automaton 
    \[
        \nfa_\gwrite{g} = \tuple{\states_\gwrite{g}, \readchars,
        \transitions_\gwrite{g}, \finals_\gwrite{g}, q^\gwrite{g}_0},
    \]
    we define the \PDS $\parampds = \tuple{\controls, \salphabet, \pdsrules,
    q_0, \finals}$ where, if $\galphabet = \set{g_0,\ldots,g_m}$, then
    \begin{itemize}
        \item $\controls = \controls_\master \times \states_\gwrite{g_0} \times
              \cdots \times \states_\gwrite{g_m} \times \brac{\galphabet \cup
              \set{\killg}}$,

        \item $q_0 = \tuple{q^\master_0, q^\gwrite{g_0}_0, \ldots,
              q^\gwrite{g_m}_0, g_0}$, 

        \item $\finals = \set{f} \times \states_\gwrite{g_0} \times \cdots
              \times \states_\gwrite{g_m} \times \brac{\galphabet \cup
              \set{\killg}}$,
    \end{itemize}
    and $\pdsrules$ is the smallest set containing all $\pdsrule{q}{a}{q'}{w}$
    where $q = \tuple{q_\master, q_0, \ldots, q_m, g}$ and,
    \begin{itemize}
        \item $q' = \tuple{q'_\master, q_0, \ldots, q_m, g}$ and
              $\pdsrule{q_\master}{a}{q'_\master}{w} \in \pdsrules_\master$, or

        \item $q' = \tuple{q'_\master, q_0, \ldots, q_m, g}$ and
              $\pdsruler{q_\master}{a}{g}{q'_\master}{w} \in \pdsrules_\master$,
              or

        \item $q' = \tuple{q'_\master, q_0, \ldots, q_m, g'}$ and
              $\pdsrulew{q_\master}{a}{g'}{q'_\master}{w} \in
              \pdsrules_\master$, or

        \item $q' = \tuple{q_\master, q_0, \ldots, q'_i, \ldots, q_m, g}$ and
              $\transition{q_i}{\gread{g}}{q'_i} \in \transitions_i$, $q_i
              \notin \finals_i$ and $w = a$, or

        \item $q' = \tuple{q_\master, q_0, \ldots, q'_i, \ldots, q_m, \killg}$
              and $\transition{q_i}{\killg}{q'_i} \in \transitions_i$, $q_i
              \notin \finals_i$ and $w = a$, or

        \item $q' = \tuple{q_\master, q_0, \ldots, q_m, g_i}$, $q_i \in
              \finals_i$ and $w = a$.
    \end{itemize}
\end{definition}
The last transition in the above definition --- which corresponds to some copy
of $\slave$ writing $g_i$ to the global store --- can be applied any number of
times; each application corresponds to a different copy of $\slave$, and, since
we are considering the parameterised problem, we can choose as many copies of
$\slave$ as are required.

\begin{lemma} \label{lma:correctness}
     The \PDS $\parampds$ has a run to some control-state in $\finals$ iff the
     parameterised reachability problem for $\master$, $\slave$, $\galphabet$,
     $g_0$ and $q$ has a positive solution.
\end{lemma}

The full proof of correctness is given in the appendix.  To construct a run
reaching $q$ from an accepting run of $\parampds$ we first observe that
$\master$ is modelled directly.  We then add a copy of $\slave$ for every
individual write to the global component of $\parampds$.  These slaves are able
to read from/write to the global component finally enabling them to perform
their designated write.  This is because (a part of) the changes to the global
store is in the read language of the required write.  

\forjade{Concerning the counter-directional, we architecturalise}{In the other
direction, we build} an accepting run of $\parampds$ from a run of the
parameterised system reaching $q$.  To this end, we observe again that we can
simulate $\master$ directly.  To simulate the slaves, we take, for every
character $g \in \galphabet$ written to the store, the copy of $\slave$
responsible for its first write.  From this we get runs of the $\nfa_\gwrite{g}$
that can be interleaved with the simulation of $\master$ and each other to
create the required accepting run, where additional writes of each $g$ are
possible by virtue of $\nfa_\gwrite{g}$ having reached an accepting state (hence
we require no further simulation for these writes).

\sideheading{Example} Let $\master$ perform the actions
$\gread{1}\gread{2}\gwrite{ok}\gread{f}$ and $\slave$ run either
$\gwrite{1}\gread{ok}\gwrite{go}$ or $\gwrite{2}\gread{go}\gwrite{f}$.  Let
$\lang_1,\ldots,\lang_4$ denote the following read languages.
\[
    \lang_\gwrite{1} = \lang_\gwrite{2} = R^\ast 
    \quad  
    \lang_\gwrite{go} = R^\ast \killg R^\ast \gread{ok} R^\ast  
    \quad
    \lang_\gwrite{f} = R^\ast \killg R^\ast \gread{go} R^\ast \\
\]
Take two slaves $\slave_1$ and $\slave_2$ and the 
run (the subscript denotes the active process):
\[
    \gwrite{1}_{\slave_1} \gread{1}_\master \gwrite{2}_{\slave_2}
    \gread{2}_\master \gwrite{ok}_\master \gread{ok}_{\slave_1}
    \gwrite{go}_{\slave_1} \gread{go}_{\slave_2} \gwrite{f}_{\slave_2}
    \gread{f}_\master \ .
\]
This can be simulated by the following actions on the global component of
$\parampds$:
\[
    \gwrite{\killg}_{\lang_3} \gwrite{1}_{\lang_1} \gread{1}_\master
    \gwrite{\killg}_{\lang_4} \gwrite{2}_{\lang_2}
    \gread{2}_\master \gwrite{ok}_\master \gread{ok}_{\lang_3}
    \gwrite{go}_{\lang_3} \gread{go}_{\lang_4}
    \gwrite{f}_{\lang_4} \gread{f}_\master \ .
\]
Note, we have scheduled the $\gwrite{\killg}$ actions immediately before the write they
correspond to.

\subsection{Complexity and Multiple Stores}

We obtain for each $g \in \galphabet$ an automaton $\nfa_\gwrite{g}$ of size
$\order{2^{2^{f(n)}}}$ in $\order{2^{2^{f(n)}}}$ time for some polynomial $f$
(using \reflma{lma:cfg2nfa}) where $n$ is the size of the problem description.
The pushdown system $\parampds$, then, has $\order{2^{2^{f'(n)}}}$ many control
states for a polynomial $f'$.  It is well known that reachability/emptiness for
\PDSs is polynomial in the size of the system (e.g.
Bouajjani\etal~\cite{BEM97}), and hence the entire algorithm takes
doubly-exponential time.  For the lower bound, one can reduce from SAT to obtain
an NP-hardness result (as shown in the appendix).  Further work is needed to
pinpoint the complexity precisely.

The algorithm presented above only applies to a single shared variable.  A more
natural model has multiple shared variables.  We may allow $k$ variables with
the addition of $k$ global components $\galphabet_1, \ldots, \galphabet_k$.  The
main change required is the use of symbols $\killg_1,\ldots, \killg_k$ rather
than simply $\killg$ and to build $\parampds$ to be sensitive to which store is
being written to (or erased with some $\killg_i$).  This does not increase the
complexity since $n = \asize{\galphabet_1} + \cdots + \asize{\galphabet_k}$ in
the above analysis and the cost of the $k$-product of variables does not exceed
the cost of the product of the $\nfa_\gwrite{g}$.  We give the full details in
the appendix.  Note that, using the global stores, we can easily encode a PSPACE
Turing machine using $\master$, without stack, and an empty $\slave$.  Hence the
problem for multiple variables is at least PSPACE-hard.

\section{Non-parameterized Reachability} \label{sec:undecidability}

We consider the reachability problem when the number of processes $n$ is fixed.
In the case when $1 \leq n \leq 2$, undecidability is clear: even with
non-atomic read/writes, the two processes can organise themselves to overcome
non-atomicity.  When $n > 2$, it becomes harder to co-ordinate the copies of
$\slave$.  A simple trick recovers undecidability.  More formally, then:
\begin{definition}[Non-parameterized Reachability]
    For given $n$ and $\naPDSs$ $\master$ and $\slave$ over $\galphabet$,
    initial store value $g_0$ and control state $q$, the non-parameterised
    reachability problem asks whether the \NPDS $\npds_n =
    \tuple{\master,\underbrace{\slave,\ldots,\slave}_{n},\galphabet, g_0}$ has a
    run to some configuration containing the control state $q$.    
\end{definition}

\begin{theorem}
    The non-parameterized reachability problem is undecidable when $n \geq 1$.
    When $n > 1$, the result holds even when $\master$ is null.
\end{theorem}
\begin{proof}
    We reduce from the undecidability of the emptiness of the intersection of
    two context-free languages.  First fix some $n \geq 2$ and two pushdown
    systems $\pds_1$, $\pds_2$ accepting the two languages $\lang_1$ and
    $\lang_2$.

    We define $\slave$ to be the disjunction of $\slave_1, \ldots, \slave_n$.
    That is, $\slave$ makes a non-deterministic choice of which $\slave_i$ to
    run ($1 \leq i \leq n$).  Let $1,\ldots,n, f, !$ be characters not in the
    alphabet of $\lang_1$ and $\lang_2$.  The process $\slave_1$ will execute,
    for each $\gamma_1\ldots \gamma_z \in \lang_1$, a sequence
    \[
        \gwrite{1}\gread{n}\gwrite{\gamma_1}\gread{!}\gwrite{\gamma_2}\gread{!}\ldots\gwrite{\gamma_z}\gread{!}\gwrite{f}
        \ .
    \]
    It is straightforward to build $\slave_1$ from $\pds_1$.  Similarly, the
    process $\slave_2$ will execute, for each $a_1\ldots a_m \in \lang_2$, a
    sequence
    \[
        \gread{1}\gwrite{2}\gread{\gamma_1}\gwrite{!}\gread{\gamma_2}\gwrite{!}\ldots\gread{\gamma_z}\gwrite{!}\gread{f}
    \]
    and move to a fresh control-state $q_f$.  It is straightforward to build
    $\slave_2$ from $\pds_2$.  The remaining processes for $3 \leq i \leq n$
    simply perform the sequence $\gread{i-1}\gwrite{i}$.

    The control-state $q_f$ can be reached iff the intersection of $\lang_1$ and
    $\lang_2$ is non-empty.  To see this, first consider a word witnessing the
    non-emptiness of the intersection.  There is immediately a run of $\npds_n$
    reaching $q_f$ where each $i$th $\slave$ process behaves as $\slave_i$.  
    
    In the other direction, take a run of $\npds_n$ reaching $q_f$.  First,
    observe that for each $1 \leq i \leq n$ there must be some copy of $\slave$
    running $\slave_i$.  This is because, otherwise, there is some $i$ not
    written to the global store, and hence all $i' \geq i$, including $n$, are
    not written.  Then $\slave_1$ can never write $f$ and $\slave_2$ can never
    move to $q_f$.  Finally, take the sequence $a_1\ldots a_m$ written by
    $\slave_1$ (and read by $\slave_2$).  This word witnesses non-emptiness as
    required.

    In the case when $n = 1$, we simply have $\master$ run $\slave_1$ and
    $\slave$ run $\slave_2$.
\end{proof}

\section{Making Ehrenfeucht and Rozenberg Constructive} 
\label{sec:cfg2reg}

We show how to make \refthm{thm:very-degenerate} constructive.  To prove
regularity, Ehrenfeucht and Rozenberg assign to each word a set of types
$\spinetype{w}$, and prove that, if $\spinetype{w} = \spinetype{w'}$, then $w \sim w'$ in
the sense of Myhill and Nerode~\cite{HU79}.  We first show how to decide
$\spinetype{w} = \spinetype{w'}$, and then show how to build the automaton.  For
the sake of brevity, we will assume familiarity with context-free grammars
(\CFGs) and their related concepts~\cite{HU79}.  

For our purposes, we consider a context-free grammar (in Chomsky normal form)
$\cfg$ to be a collection of rules of the form $A \prodarr BC$ or $A \prodarr
a$, where $A, B$ and $C$ are \emph{non-terminals} and $a$ is a \emph{terminal}
in $\oalphabet$.  There is also a designated \emph{start} non-terminal $S$.  A
word $w$ is in $\langof{\cfg}$ if there is a \emph{derivation-tree} with root
labelled by $S$ such that an internal node labelled by $A$ has left- and
right-children labelled by $B$ and $C$ when we have $A \prodarr BC$ in the
grammar and a leaf node is labelled by $a$ when it has parent labelled by $A$
(with one child) and $A \prodarr a$ is in the grammar.  Furthermore $w$ is the
\emph{yield} of the tree; that is, $w$ labels the leaves.  Note, all nodes must
be labelled according to the scheme just described.  One can also consider the
derivation of $w$ in terms of \emph{rewrites} from $S$, where the parent-child
relationship in the tree gives the requires rewriting steps.

\subsection{Preliminaries}

We first recall some relevant definitions from Ehrenfeucht and Rozenberg.  We
write $\numchar{a}{w}$ to mean the number of occurrences of the character $a$ in
the word $w$.
\begin{definition}[Type of a Word]
    Let $\oalphabet$ be an alphabet and let $x, w \in \oalphabet^\ast$, We say
    that $w$ is of type $x$, or that $x$ is a type of $w$ (denoted
    $\wordtype{x}{w}$) if
    \begin{enumerate}
        \item for every $a \in \oalphabet$, $\numchar{a}{x} \leq 1$, and

        \item there exists a homomorphism $h$ such that
        \begin{enumerate}
            \item for every $a \in \oalphabet$, $h(a) \in a \cup a
                  \oalphabet^\ast a$, and

            \item $h(x) = w$.
        \end{enumerate}
    \end{enumerate}
    If $x$ satisfies the above, we also say that $x$ is a type in
    $\oalphabet^\ast$.
\end{definition}

Given a \CFG $\cfg$ in Chomsky normal form, we assume a derivation tree $\dtree$
of $\cfg$ is a labelled tree where all internal nodes are labelled with the
non-terminal represented by the node, and all leaf nodes are labelled by their
corresponding characters in $\oalphabet$.  Given a derivation tree $\dtree$,
Ehrenfeucht and Rozenberg define a marked tree $\marked{\dtree}$ with an
expanded set of non-terminals and terminals.  Simultaneously, we will define the
spine of a marked tree.  Intuitively, we take a path in the tree and mark it
with the productions of $\cfg$ that have been used and the directions taken.

Given an alphabet of terminals and non-terminals $\lalphabet$ and a derivation
tree $\dtree$, \longshort{define the alphabet}{let} $\marked{\lalphabet} =
\setcomp{(A, B, C, k)}{k \in \set{1,2} \land A \prodarr BC \in \cfg} \cup
\setcomp{(A, a)}{A \prodarr a \in \cfg}$.  This is the marking alphabet of
$\cfg$.  

\begin{definition}[Spine of a Derivation Tree]
    Let $\dtree$ be a derivation tree in $\cfg$ and let $\dpath = v_0\ldots v_s$
    be a path in $\dtree$ where $s \geq 1$, $v_0$ is the root of $\dtree$, $v_s$
    is a leaf of $\dtree$ and $\dlab{v_0},\ldots,\dlab{v_s}$ are the labels
    corresponding to nodes of $\dpath$.  Now for each node $v_j$, $0 \leq j \leq
    s$, change its label to $\dmarklab{v_j}$ as follows:
    \begin{enumerate}
        \item if $A \prodarr BC$ is the production used to rewrite the node $j$
              (hence $\dlab{v_j} = A$) and $v_j$ has a direct descendant to the
              left of $\dpath$, then $\dlab{v_j}$ is changed to $\dmarklab{v_j}
              = (A, B, C, 1)$,

        \item if $A \prodarr BC$ is the production used to rewrite the node $j$
              and $v_j$ has a direct descendant to the right of $\dpath$, then
              $\dlab{v_j}$ is changed to $\dmarklab{v_j} = (A, B, C, 2)$,

        \item if $A \prodarr a$ is the production used to rewrite the node $j$
              then $\dlab{v_j}$ is changed to $\dmarklab{v_j} = (A, a)$,

        \item $\dmarklab{v_s} = \dlab{v_s}$.
    \end{enumerate}

    The resulting tree is called the \emph{marked $\dpath$-version} of $\dtree$
    and denoted by $\markedtree{\dpath}{\dtree}$.  The word
    $\dmarklab{v_0}\ldots\dmarklab{v_s}$ is referred to as the \emph{spine} of
    $\markedtree{\dpath}{\dtree}$ and denoted by
    $\spine{\markedtree{\dpath}{\dtree}}$.
\end{definition}

We write $\hasspine{w}{z}$ whenever there exists some $u$ such that the word
$wu$ has a derivation tree $\dtree$ in $\cfg$ with a path $\dpath$ ending on the
last character of $w$ and with $\spine{\markedtree{\dpath}{\dtree}} = z$.  Then,
we have $\spinetype{w} = \setcomp{x}{\hasspine{w}{z} \land \wordtype{x}{z}}$.
Intuitively, this is the \emph{spine-type} of $w$.

Finally, Ehrenfeucht and Rozenberg show that, whenever all strong iterative
pairs of $\cfg$ are very degenerate, then $\spinetype{w} = \spinetype{w'}$
implies $w \sim w'$.  Since there are a finite number of types $x$, we have
regularity by Myhill and Nerode.

\subsection{Building the Automaton}

We show how to make the above result constructive.  The first step is to decide
$\spinetype{w} = \spinetype{w'}$ for given $w$ and $w'$.  To do this, from
$\cfg$ and some type $x$, we build $\typedg{\cfg}{x}$ which generates all $w$
such that $\hasspine{w}{z}$ holds for some $z$ of type $x$.  Thus $x \in
\spinetype{w}$ iff $w \in \langof{\typedg{\cfg}{x}}$. 

First note that there is a simple (polynomial) regular automaton $\typenfa{x}$
recognising, for $x = a_1 \ldots a_s$ the language 
\[ 
    \brac{a_1 \cup a_1\marked{\lalphabet}^\ast a_1}\ldots\brac{a_s \cup
    a_s\marked{\lalphabet}^\ast a_s} 
\] 
and $z \in \langof{\typenfa{x}}$ iff $z$ is of type $x$.  The idea is to build
this automaton into the productions of $\cfg$ to obtain $\typedg{\cfg}{x}$ such
that all characters to the left (inclusive) of the path chosen by $\typenfa{x}$
are kept, while all those to the right are erased.  

\begin{definition}[$\typedg{\cfg}{x}$] 
    For a given word type $x$ and \CFG $\cfg$, the grammar $\typedg{\cfg}{x}$
    has the following production rules:
    \begin{itemize}
        \item all productions in $\cfg$,

        \item $A_q \prodarr B_{q'} C_\varepsilon$ for each $A \prodarr BC \in
              \cfg$ and $\transition{q}{(A, B, C, 1)}{q'}$ in $\typenfa{x}$,

        \item $A_q \prodarr B C_{q'}$ for each $A \prodarr BC \in \cfg$ and
              $\transition{q}{(A, B, C, 2)}{q'}$ in $\typenfa{x}$,

        \item $A_q \prodarr a$ for each $A \prodarr a \in \cfg$ and
              $\transition{q}{(A, a)}{q'}$ in $\typenfa{x}$ where $q'$ is a final
              state,

        \item $A_\varepsilon \prodarr B_\varepsilon C_\varepsilon$ for each $A
              \prodarr BC \in \cfg$,

        \item $A_\varepsilon \prodarr \varepsilon$ for each $A \prodarr a \in
              \cfg$.
    \end{itemize}
    The initial non-terminal is $S_{q_0}$ where $S$ is the initial non-terminal
    of $\cfg$ and $q_0$ is the initial state of $\typenfa{x}$.  
\end{definition}

The correctness of $\typedg{\cfg}{x}$ is straightforward and hence relegated to
the appendix.

\begin{lemma} \label{lma:typed-cfg}
    For all $w$, we have $w \in \langof{\typedg{\cfg}{x}}$ iff $x \in
    \spinetype{w}$.
\end{lemma}

\begin{lemma}[Deciding $\spinetype{w} = \spinetype{w'}$] \label{lma:check-eq}
    For given $w$ and $w'$, we can decide $\spinetype{w} = \spinetype{w'}$ in
    $\order{2^{f(n)}}$ time for some polynomial $f$ where $n$ is the size of
    $\cfg$.
\end{lemma}
\begin{proof}
    For a given alphabet $\marked{\lalphabet}$, there are $\sum^{m}_{r = 1} r!$
    types where $m = \asize{\marked{\lalphabet}}$.  Since $m$ is polynomial in
    $n$, there are $\order{2^{f(n)}}$ word types.  Hence, we simply check $w \in
    \langof{\typedg{\cfg}{x}}$ and $w' \in \langof{\typedg{\cfg}{x}}$ for each
    type $x$.  This is polynomial for each $x$, giving $\order{2^{f(n)}}$ in
    total.
\end{proof}

From this, we can construct, following Myhill and Nerode, the required
automaton, using a kind of fixed point construction beginning with an automaton
containing the state $q_\varepsilon$ from which the equivalence class associated
to the empty word will be accepted.

\begin{lemma} \label{lma:cfg2nfa}
    For a \CFG $\cfg$ such that all strong iterative pairs are very degenerate,
    we can build an \NFA $\nfa$ of $\order{2^{2^{f(n)}}}$ size in the same
    amount of time, where $n$ is the size of $\cfg$.
\end{lemma}
\begin{proof}
    Let $\cfg$ be a \CFG such that all strong iterative pairs are degenerate.  We
    build an \NFA $\nfa$ such that $\langof{\cfg} = \langof{\nfa}$ by the following
    worklist algorithm.

    \begin{enumerate}
        \item Let the worklist contain only $\varepsilon$ (the empty word) and
              $\nfa$ have the initial state $q_\varepsilon$.

        \item \label{alg:cfg2reg:loop} Take a word $w$ from the worklist.

        \item If $w \in \langof{\cfg}$, make $q_w$ a final state.

        \item For each $a \in \oalphabet$
              \begin{enumerate}
                  \item if there is no state $q_{w'}$ such that $\spinetype{wa} =
                        \spinetype{w'}$, add $q_{wa}$ to $\nfa$ and add $wa$ to the
                        worklist,

                  \item take $q_{w'}$ in $\nfa$ such that $\spinetype{wa} =
                        \spinetype{w'}$,

                  \item add the transition $\transition{q_w}{a}{q_{w'}}$ to $\nfa$.
              \end{enumerate}
        
        \item If the worklist is not empty, go to point~\ref{alg:cfg2reg:loop},
              else, return $\nfa$.
    \end{enumerate}

    Since this follows the Myhill-Nerode construction, using $\spinetype{w} =
    \spinetype{w'}$ as a proxy for $w \sim w'$, we have that the algorithm
    terminates and is correct.  Hence, with the observation that there are
    $\order{2^{2^{f(n)}}}$ different values of the sets $\spinetype{w}$, we have
    the lemma.
\end{proof}

\section{Conclusions and Future Work} \label{sec:conclusion}

In this work, we have studied the parameterised master/slave reachability
problem for pushdown systems with a global store.  This provides an extension of
work by Kahlon which did not allow a master process, and communication was via
anonymous synchronisation; however, this is obtained at the expense of atomic
accesses to global variables.  Our algorithm introduces new techniques to
pushdown system analysis.

An initial inspiration for this work was the study of weak-memory models, which
do not guarantee that --- in a multi-threaded environment --- memory accesses
are \emph{sequentially consistent}.  In general, if atomic read/writes are
permitted, the verification problem is harder (for example, Atig\etal relate the
finite-state case to lossy channel machines~\cite{ABBM10}); hence, we removed
atomicity as a natural first step.  It is not clear how to extend our algorithm
to accommodate weak-memory models and it remains an interesting avenue of future
work.

Another concern is the complexity gap between the upper and lower bounds.  We
conjecture that the upper bound can be improved, although we may require a new
approach, since the complexity comes from the construction of regular read
languages.  A related question is whether we can improve the size of the
automata $\nfa_\gwrite{g}$.  Since a \PDS of size $n$ can recognise the language
$\set{a^{2^n}}$, we have a read language requiring an exponential number of $a$
characters; hence, the $\nfa_\gwrite{g}$ must be at least exponential in the
worst case.  It is worth noting that Meyer and Fischer give a language whose
\emph{deterministic} regular automaton is doubly-exponential in the size of the
corresponding deterministic \PDS~\cite{MF71}.  However, in the appendix, we
provide an example showing that this language is not very degenerate.  If the
\PDS is not deterministic, Meyer and Fischer prove there is no bound, in
general, on the relationship in sizes.

Finally, we may also consider applications to recursive ping-pong protocols in
the spirit of Delzanno\etal~\cite{DES06}.

\sideheading{Acknowledgments} Nous remercions Jade Alglave pour plusieurs
discussions qui ont amorc\'ees ce travail.  This work was funded by EPSRC grant
EP/F036361/1.  We also thank the anonymous reviewers and Ahmed Bouajjani for
their helpful remarks.

    \longshort{
        \bibliographystyle{plain} 
        \bibliography{references}

        \newpage \appendix

\section{Proofs for \refsec{sec:result}}

The proof of \reflma{lma:correctness} is split into the following two lemmas.

\begin{lemma}
    The \PDS $\parampds$ has a run to some control-state in $\finals$, then the
    parameterised reachability problem for $\master$, $\slave$, $\galphabet$,
    $g_0$ and $q$ has a positive solution.
\end{lemma}
\begin{proof}
Take an accepting run of $\parampds$.  We can extract a number of sequences
from this run.  First, let $G = g^1_G,\ldots, g^z_G$ be the sequence of
values written to the global (last) component of $\parampds$'s
control-state.  Note, $g^1_G = g_0$.  Then, for each $g \in \galphabet$ that
is written to the global component, let $R_g$ be the sequence of read and
$\killg$ events that took $\nfa_\gwrite{g}$ from $q^\gwrite{g}_0$ to a state
in $\finals_\gwrite{g}$.  Since this is accepted by the read language of
$g$, there is a subword $\gread{g^1},\ldots,\gread{g^{x}}$ of $R_g$ and
sequences of writes $W_0,\ldots,W_x$ such that $W_0 \gread{g^1} W_1 \ldots
\gread{g^x} W_x \gwrite{g}$ is a run of $\slave$ (with internal transitions
hidden).

Furthermore, let $\killg^i$ be a sequence of $\killg$ characters the same
length as $W_i$.  Notice, we can fix a sub-sequence $G_g = \killg^0 g^1
\cdots g^x \killg^x g$ of $R_g$ corresponding to a run of $\slave$ in the
sense that, $\killg$ characters represent some write action, the $g^h$ for
all $1 \leq h \leq x$ are read events of $g^h$, and $g$ is a write of $g$.
Similarly, $\master$ has a sub-sequence $G_\master$ leading to $q$.  This
sequence is mapped on to $G$ as follows.  The sequence $G$ partitions the
run of $\parampds$ into contiguous sections with each $g^i_G$ beginning a
new section.  Since $G_g$ is a sub-sequence of $R_g$ which is in turn a
sub-sequence of the run of $\parampds$, there is a natural mapping of
elements of $G_g$ to the transitions in the run of $\parampds$.  Each
character is mapped to the element of $G$ that begins the section the
transition occurs in.  Similarly, $\master$ has a sequence $G_\master$
leading to $q$.

We create the \NPDS which has a unique process $\slave$ for each $g^i_G$ in
$G$ that is not $\killg$ and is not written by $\master$ (that is, a process
for each individual write).  We build the run in $z$ segments: one for each
$g^i_G$.  In each segment, all processes whose sub-sequence $G_g$ or
$G_\master$ maps a character onto $g^i_G$ will be scheduled to make the
corresponding transitions.  These can be scheduled in any order, except the
process running first in the segment must be the process responsible for
writing $g^i_G$.  When $g^i_G = \killg$, the process will not write $\killg$
to the store, but some other character.  Since no process reads $\killg$
this is safe.

Observe that there may be some $g^i_G$ that are not written by any process.
In this case $g^i_G = \killg$ (since we allowed $\killg$ to occur at any
time) and, because no process reads $\killg$, the corresponding segment is
merely $\varepsilon$.
\end{proof}

\begin{lemma}
If the parameterised reachability problem for $\master$, $\slave$,
$\galphabet$, $g_0$ and $q$ has a positive solution, then $\parampds$ has a
run to some control-state in $\finals$.
\end{lemma}
\begin{proof}
Take a run $C = c_0 c_1 \ldots c_z$ of the \NPDS with $n$ copies of $\slave$
that reaches $q$.  From this, we build an accepting run $\pi$ of
$\parampds$.  The initial configuration of $\pi$ is $\tuple{q^\master_0,
q^\gwrite{g_0}_0, \ldots, q^\gwrite{g_m}_0, g_0}$.  Assume we have a run
$\pi_i$ corresponding to the run of the \NPDS up to $c_i$.  This run will
have the property that the first component (the control-state of $\master$)
of the last configuration in $\pi_i$ will match the control-state of
$\master$ in $c_i$.  Hence, $\pi_z$ will be the required accepting run.  

Take the first write of $\gwrite{g}$ of each $g \in \galphabet$ that is
written by some copy of $\slave$.  Take the run of $\slave$ that produced
the write which is a sequence of reads and writes $W_0 R_1 W_1 \ldots R_x
W_x$ (with internal moves omitted).  Let $\killg_j$ be a sequence of
$\killg$ characters with the same length as $W_j$.  There is as accepting
run $q^\gwrite{g}_0 \runtransition{\gamma_1} q^\gwrite{g}_1
\runtransition{\gamma_2} \cdots \runtransition{\gamma_{y}} q^\gwrite{g}_{y}$
of $\nfa_\gwrite{g}$ where $\killg_0 R_1 \killg_1 R_1 \ldots R_x \killg_x =
\gamma_1\ldots\gamma_{y}$.  Furthermore, $\gamma_1\ldots\gamma_{y}$ can
be mapped onto a sub-word of the sequence of actions taken on the global
component up to the first write of $g$.

Let $\tuple{q^\master_{i_\master}, q^\gwrite{g_0}_{i_{g_0}}, \ldots,
q^\gwrite{g_m}_{i_{g_m}}, g^i}$ be the final configuration of $\pi_i$.  We
extend $\pi_i$ with the following transitions, in order of appearance.
\begin{itemize}
    \item For all $g$ such that we have a maximal path $q^\gwrite{g}_{i_g}
          \runtransition{\gread{g}} \cdots \runtransition{\gread{g}}
          q^\gwrite{g}_{i_g + 1}$, make the transitions to
          $q^\gwrite{g}_{i_g + 1}$.  (That is, read $g$ as many times as
          possible.)

    \item If the transition between $c_i$ and $c_{i+1}$ is a move of
          $\master$, then simulate the move directly.
    
    \item If the transition is a write move $\gwrite{g}$ by a copy of
          $\slave$ which is not responsible for the first write of $g$, but
          is responsible for for the first write of some other $g'$, then
          advance $q^\gwrite{g'}_{i_{g'}} \runtransition{\killg}
          q^\gwrite{g'}_{i_{g'}}$, setting the global component to $\killg$
          as required.  Note that the transition from
          $q^\gwrite{g'}_{i_{g'}}$ must be a $\killg$ move since it is a
          write move of $\slave$ and all preceding reads and writes have
          been simulated.

    \item Further to the above, if it is a write of $g$ by some $\slave$, we
          know that $q^\gwrite{g}_{i_{g}}$ is an accepting state of
          $\nfa_\gwrite{g}$.  This is because we have been simulating the
          sequence $W_0 R_1 W_1 \ldots R_x W_x$ with the accepting run
          $\killg_0 R_1 \killg_1 \ldots R_x \killg_x$.  Hence we can (and
          do) perform the write of $g$ to the global component.

    \item Other types of transitions have no further updates to $\pi_i$.  In
          particular, if the transition is a read move by some copy of
          $\slave$ we do not add any transitions (these moves are taken care
          of more eagerly above).
\end{itemize}
This completes the construction of $\pi_i$, and thus $\pi_y$ gives us a
required accepting run of $\parampds$.
\end{proof}

\section{Non-Atomic Pushdown Systems with Multiple Variables}
        \label{sec:multi-vars}

\subsection{Model Definition}

\begin{definition}[Non-atomic Pushdown Systems with Multiple-Variables]
    Over a partitioned finite alphabet $\galphabet = \galphabet_1 \uplus \cdots
    \uplus \galphabet_k$, a \emph{non-atomic pushdown system} (\naPDS) is a
    tuple $\pds = \tuple{\controls, \salphabet, \pdsrules, q_0, \galphabet_1,
    \ldots, \galphabet_k}$ where $\controls$ is a finite set of control-states,
    $\salphabet$ is a finite stack alphabet, $q_0 \in \controls$ is a designated
    initial control state and $\pdsrules \subseteq \brac{\controls \times
    \salphabet} \times \brac{\gread{\galphabet} \cup \gwrite{\galphabet} \cup
    \set{\varepsilon}} \times \brac{\controls \times \salphabet^\ast}$.
\end{definition}

\begin{definition}[Networks of \naPDSs with Multiple Variables]
    A network of $n$ \emph{non-atomic pushdown systems} (\NPDS) is a tuple
    $\npds = \tuple{\pds_1, \ldots, \pds_n, \galphabet_1, \ldots, \galphabet_k,
    g^1_0, \ldots, g^k_0}$ where, for all $1 \leq i \leq n$, $\pds_i =
    \tuple{\controls_i, \salphabet_i, \pdsrules_i, q^i_0, \galphabet_1, \ldots,
    \galphabet_k}$ is a $\NPDS$ over $\galphabet_1, \ldots, \galphabet_k$ and
    for all $1 \leq i \leq k$, $g^i_0 \in \galphabet_i$ is the initial value of
    the $i$th global store.
\end{definition}

A configuration of an \NPDS is a tuple $\config{q_1, w_1, \ldots, q_n, w_n}{g_1,
\ldots, g_k}$ where $g_i \in \galphabet_i$ for each $1 \leq i \leq k$, and for
each $1 \leq i \leq n$, $q_i \in \controls_i$ and $w_i \in \Sigma^\ast_i$.  We
have a transition 
\[ 
    \config{q_1, w_1, \ldots, q_n, w_n}{g_1,\ldots,g_k} \npdstran \config{q'_1,
    w'_1, \ldots, q'_n, w'_n}{g'_1,\ldots,g'_k} 
\] 
whenever, for some $1 \leq i \leq n$ and all $1 \leq j \leq n$ with $i \neq j$
we have $q'_j = q_j$, $w'_j = w_j$, and
\begin{itemize}
    \item $\localconfig{q_i}{w_i} \pdstran \localconfig{q'_i}{w'_i}$ is a
          transition of $\pds_i$ and for all $1 \leq l \leq k$, $g'_l = g_l$; or

    \item $\localconfig{q_i}{w_i} \pdstranr{g_l} \localconfig{q'_i}{w'_i}$ is a
          transition of $\pds_i$ for some $1 \leq l \leq k$ and for all $1 \leq
          l' \leq k$, $g'_{l'} = g_{l'}$; or

    \item $\localconfig{q_i}{w_i} \pdstranw{g'_l} \localconfig{q'_i}{w'_i}$ is a
          transition of $\pds_i$ for some $1 \leq l \leq k$ and for all $1 \leq
          l' \leq k$ such that $l' \neq l$, $g'_{l'} = g_{l'}$.
\end{itemize}
A path $\run$ of $\npds$ is a sequence of configurations $c_1 c_2 \ldots c_z$
such that, for all $1 \leq i < z$, $c_i \npdstran c_{i+1}$.  A run of $\npds$ is
a path such that $c_1 = \config{q^1_0, \sbot, \ldots, q^n_0,
\sbot}{g^1_0,\ldots,g^k_0}$.

\subsection{Reachability Analysis} 

In this section, we aim prove the following theorem.
\begin{theorem}
    The parameterised reachability problem for \NPDSs with multiple variables is
    decidable.
\end{theorem}

Again, we assume $q$ is a control-state of $\master$.  The idea is the same as
the single variable case, except for some minor adjustments to handle the extra
variables.

\subsubsection{Regular Read Languages} 

Given a non-atomic pushdown system $\pds$ we define for each $g \in \galphabet$
the pushdown system $\pds_{\gwrite{g}}$ which is $\pds$ augmented with a new
unique control-state $f$, and a transition $\pdsrule{q}{a}{f}{a}$ whenever
$\pds$ has a rule $\pdsrulew{q}{a}{g}{q'}{w}$.  Furthermore, replace all
$\pdsrulew{q}{a}{g'}{q'}{w}$ rules with $\pdsrulebasic{q}{a}{\killg_i}{q'}{w}$
where $\killg_i \notin \galphabet_1 \cup \cdots \cup \galphabet_k$ and $g' \in
\galphabet_i$. 

Again, we interpret $f$ as the sole accepting control state of
$\pds_{\gwrite{g}}$ giving the read language $\lang_\gwrite{g}$ for $\gwrite{g}$
defined as 
\[
    \lang_\gwrite{g} = \setcomp{\readchars^\ast\gamma_1\readchars^\ast\ldots
    \readchars^\ast\gamma_z\readchars^\ast}{\gamma_1\ldots\gamma_z \in
    \langof{\pds_\gwrite{g}}}
\]
where $\readchars = \setcomp{\gread{g'}}{g' \in \galphabet} \cup
\set{\killg_1,\ldots,\killg_k}$.

\begin{lemma}
    For all $g \in \galphabet$, $\lang_\gwrite{g}$ is regular and an \NFA $\nfa$
    accepting $\lang_\gwrite{g}$, of doubly-exponential size, can be constructed
    in doubly-exponential time.    
\end{lemma}
\begin{proof}
    Identical to the single variable case.
\end{proof}

\subsubsection{Simulating the System}

We build a PDS that recognises a non-empty language iff the parameterised
reachability problem has a positive solution.  The intuition behind the
construction of $\parampds$ is the same as the single variable case, except
minor adjustments are needed to handle the interaction with multiple variables.

\begin{definition}[$\parampds$]
    Given an \naPDS $\master = \tuple{\controls_\master, \salphabet,
    \pdsrules_\master, q^\master_0, \galphabet_1, \ldots, \galphabet_k}$ over
    $\galphabet = \galphabet_1 \uplus \cdots \uplus \galphabet_k$ with initial
    values $g^1_0,\ldots,g^k_0$, a control-state $f \in \controls_\master$, and,
    for each $g \in \galphabet$, a regular automaton $\nfa_\gwrite{g} =
    \tuple{\states_\gwrite{g}, \readchars, \transitions_\gwrite{g},
    \finals_\gwrite{g}, q^\gwrite{g}_0}$, we define the \PDS $\parampds =
    \tuple{\controls, \salphabet, \pdsrules, q_0, \finals}$ where
    \begin{itemize}
        \item we let, for all $i$, $\galphabet_i = \set{g^i_0,\ldots,g^i_{m_i}}$
              and, 

        \item let $\vec{\states} = \states_\gwrite{g^1_0} \times \cdots \times
              \states_\gwrite{g^1_{m_1}} \times \cdots \times
              \states_\gwrite{g^k_0} \times \cdots \times
              \states_\gwrite{g^k_{m_k}}$, then

        \item $\controls = \controls_\master \times \vec{\states} \times
              \brac{\galphabet_1 \cup \set{\killg_1}} \times \cdots \times
              \brac{\galphabet_k \cup \set{\killg_k}}$,

        \item $q_0 = \tuple{q^\master_0, q^\gwrite{g^1_0}_0, \ldots,
              q^\gwrite{g^k_{m_k}}_0, g^1_0,\ldots,g^k_0}$, 

        \item $\finals = \set{f} \times \vec{\states} \times \brac{\galphabet_1
              \cup \set{\killg_1}} \times \cdots \times \brac{\galphabet_k \cup
              \set{\killg_k}}$,
    \end{itemize}
    and $\pdsrules$ is the smallest set containing all $\pdsrule{q}{a}{q'}{w}$
    where $q = \tuple{q_\master, q^1_0, \ldots, q^k_{m_k}, g_1,\ldots,g_k}$ and,
    \begin{itemize}
        \item $q' = \tuple{q'_\master, q^1_0, \ldots, q^k_{m_k}, g_1,\ldots,
              g_k}$ and $\pdsrule{q_\master}{a}{q'_\master}{w} \in
              \pdsrules_\master$, or

        \item $q' = \tuple{q'_\master, q^1_0, \ldots, q^k_{m_k}, g_1,\ldots,
              g_k}$ and $\pdsruler{q_\master}{a}{g_i}{q'_\master}{w} \in
              \pdsrules_\master$ for some $i$, or

        \item $q' = \tuple{q'_\master, q^1_0, \ldots, q^k_{m_k},
              g_1,\ldots,g'_i,\ldots,g_k}$ and
              $\pdsrulew{q_\master}{a}{g'_i}{q'_\master}{w} \in
              \pdsrules_\master$ for some $g'_i \in \galphabet_i$, or

        \item $q' = \tuple{q_\master, q^1_0, \ldots, p^i_j, \ldots, q^k_{m_k},
              g_1,\ldots,g_k}$ and $\transition{q^i_j}{\gread{g_l}}{p^i_j} \in
              \transitions^i_j$ for some $l$, $q^i_j \notin \finals^i_j$ and $w
              = a$, or

        \item $q' = \tuple{q_\master, q^1_0, \ldots, p^i_j, \ldots, q^k_{m_k},
              g_1,\ldots,\killg_l,\ldots,g_k}$ and
              $\transition{q^i_j}{\killg_l}{p^i_j} \in \transitions^i_j$, $q^i_j
              \notin \finals^i_j$ and $w = a$, or

        \item $q' = \tuple{q_\master, q^1_0, \ldots, q^k_{m_k}, g_1, \ldots,
              g^i_j, \ldots, g_k}$, $q^i_j \in \finals^i_j$ and $w = a$.
    \end{itemize}
\end{definition}

We have the following property.
\begin{lemma}
    The \PDS $\parampds$ has a run to some control-state in $\finals$ iff the
    parameterised reachability problem for $\master$, $\slave$,
    $\galphabet_1,\ldots,\galphabet_k$, $g^1_0,\ldots,g^k_0$ and $q$ has a
    positive solution.
\end{lemma}

We prove this property in the following lemmas, and conclude that the
parameterised reachability problem with multiple variables is decidable.

\begin{lemma} \label{lem:multi-sound}
     The \PDS $\parampds$ has a run to some control-state in $\finals$, then the
     parameterised reachability problem for $\master$, $\slave$,
     $\galphabet_1,\ldots,\galphabet_k$, $g^1_0,\ldots,g^k_0$ and $q$ has a
     positive solution.
\end{lemma}
\begin{proof}
    Take an accepting run of $\parampds$.  We can extract a number of sequences
    from this run.  First, let $G = \vec{g}^1,\ldots, \vec{g}^z$ be the sequence
    of updates to the global (last $k$) components of $\parampds$'s
    control-state.  That is, $g^1 = \tuple{g^1_0, \ldots, g^k_0}$, and
    $\vec{g}^{i+1}$ is generated from $\vec{g}^i$ by the next change to a global
    component.  Then, for each $g$ that is written to a global component, let
    $R_g$ be the sequence of read and $\killg_1,\ldots,\killg_k$ events that
    took $\nfa_\gwrite{g}$ from $q^\gwrite{g}_0$ to a state in
    $\finals_\gwrite{g}$.  Since this is accepted by the read language of $g$,
    there is a subword $\gread{g^1},\ldots,\gread{g^{x}}$ of $R_g$ and sequences
    of writes $W_0,\ldots,W_x$ such that $W_0 \gread{g^1} W_1 \ldots \gread{g^x}
    W_x \gwrite{g}$ is a run of $\slave$ (with internal transitions hidden).
    
    Furthermore, let $\killg^i$ be a sequence of actions derived from $W_i$ by
    replacing each write to a variable $j$ with the character $\killg_j$.  We
    can fix a sub-sequence $G_g = \killg^0 g^1 \cdots g^x \killg^x g$ of $R_g$
    corresponding to the run of $\slave$ above.  This sequence is mapped on to
    $G$ as follows.  The sequence $G$ partitions the run of $\parampds$ into
    contiguous sections with each $\vec{g}^i$ beginning a new section.  Since
    $G_g$ is a sub-sequence of $R_g$ which is in turn a sub-sequence of the run
    of $\parampds$, there is a natural mapping of elements of $G_g$ to the
    transitions in the run of $\parampds$.  Each character is mapped to the
    element of $G$ that begins the section the transition occurs in.  Similarly,
    $\master$ has a sequence $G_\master$ leading to $q$.

    We create the \NPDS which has a unique process $\slave$ for each $\vec{g}^i$
    in $G$ that is not a $\killg_j$ event for some $j$ and is not written by
    $\master$ (that is, a process for each individual write).  We build the run
    in $z$ segments: one for each $\vec{g}^i$.  In each segment, all processes
    whose sub-sequence $G^g$ (when the update given by $\vec{g}^i$ is a write of
    the character $g$) or $G_\master$ maps a character onto $\vec{g}^i$ will be
    scheduled to make the corresponding transitions (including internal
    transitions).  These can be scheduled in any order, except the process
    running first in the segment must be the process responsible for writing
    $g$.  When $\vec{g}^i$ is a write of $\killg_j$, the process will not write
    $\killg_j$ to the $j$th component of the store, but some other character.
    Since no process reads $\killg_j$ this is safe.

    Observe that there may be some updates $\vec{g}^i$ that are not written by
    any process.  In this case the update is the write of some $\killg_j$ (since
    we allowed $\killg_j$ to occur at any time) and, because no process reads
    $\killg_j$, the corresponding segment is merely $\varepsilon$.
\end{proof}

\begin{lemma}
    If the parameterised reachability problem for $\master$, $\slave$,
    $\galphabet_1,\ldots,\galphabet_k$, $g^1_0, \ldots, g^k_0$ and $q$ has a
    positive solution, then $\parampds$ has a run to some control-state in
    $\finals$.
\end{lemma}
\begin{proof}
    Take a run $C = c_0 c_1 \ldots c_z$ of the \NPDS with $n$ copies of $\slave$
    that reaches $q$.  From this, we build an accepting run $\pi$ of
    $\parampds$.  The initial configuration of $\pi$ is 
    \[ 
        \tuple{q^\master_0, q^\gwrite{g^1_0}_0, \ldots, q^\gwrite{g^k_{m_k}}_0,
        g^1_0,\ldots,g^k_0} \ . 
    \]
    Assume we have a run $\pi_i$ corresponding to the run of the \NPDS up to
    $c_i$.  This run will have the property that the first component (the
    control-state of $\master$) of the last configuration in $\pi_i$ will match
    the control-state of $\master$ in $c_i$.  Hence, $\pi_z$ will be the
    required accepting run.  

    Take the first write of $\gwrite{g}$ for each $g \in \galphabet$ that is
    written by some copy of $\slave$.  Take the run of $\slave$ that produced
    the write which is a sequence of reads and writes $W_0 R_1 W_1 \ldots R_x
    W_x$ (with internal moves omitted).  Let $\killg^j$ be a sequence of
    $\killg_1,\ldots,\killg_k$ characters derived from $W_j$ as in the proof of
    \reflma{lem:multi-sound}.  There is an accepting run of $\nfa_\gwrite{g}$
    \[
        q^\gwrite{g}_0 \runtransition{\gamma_1} q^\gwrite{g}_1
        \runtransition{\gamma_2} \cdots \runtransition{\gamma_{y}}
        q^\gwrite{g}_{y}
    \]
    where $\killg^0 R_1 \killg^1 R_1 \ldots R_x \killg^x =
    \gamma_1\ldots\gamma_{y}$.  Furthermore, $\gamma_1\ldots\gamma_{y}$ can
    be mapped onto a sub-word of the sequence of actions taken on the global
    components up to the first write of $g$.

    Let $\tuple{q^\master_{i_\master}, q^\gwrite{g^1_0}_{i_{g^1_0}}, \ldots,
    q^\gwrite{g^k_{m_k}}_{i_{g^k_{m_k}}}, g_1,\ldots,g_k}$ be the final
    configuration of $\pi_i$.  We extend $\pi_i$ with the following transitions,
    in order of appearance.
    \begin{itemize}
        \item For all $g$ such that we have a maximal path $q^\gwrite{g}_{i_g}
              \runtransition{\gread{g^1}} \cdots \runtransition{\gread{g^y}}
              q^\gwrite{g}_{i_g + 1}$ where $g^{j}$ for $1 \leq j \leq y$ are
              characters in $\set{g_1,\ldots,g_k}$, make the transitions to
              $q^\gwrite{g}_{i_g + 1}$.  (That is, read the current global store
              as many times as possible.)

        \item If the transition between $c_i$ and $c_{i+1}$ is a move of
              $\master$, then simulate the move directly.
        
        \item If the transition is a write move $\gwrite{g}$ for some $g \in
              \galphabet$ by a copy of $\slave$ which is not responsible for the
              first write of $g$, but is responsible for for the first write of
              some other $g'$, then advance $q^\gwrite{g'}_{i_{g'}}
              \runtransition{\killg_j} q^\gwrite{g'}_{i_{g'}}$, setting the
              $j$th global component to $\killg_j$ as required.  Note that the
              transition from $q^\gwrite{g'}_{i_{g'}}$ must be a $\killg_j$ move
              since it is a write move to the $j$th component of $\slave$ and
              all preceding reads and writes have been simulated.

        \item Further to the above, if it is a write of $g$ by some $\slave$, we
              know that $q^\gwrite{g}_{i_{g}}$ is an accepting state of
              $\nfa_\gwrite{g}$.  This is because we have been simulating the
              sequence $W_0 R_1 W_1 \ldots R_x W_x$ with the accepting run
              $\killg_0 R_1 \killg_1 \ldots R_x \killg_x$.  Hence we can (and
              do) perform the write of $g$ to the global component.

        \item Other types of transitions have no further updates to $\pi_i$.  In
              particular, if the transition is a read move by some copy of
              $\slave$ we do not add any transitions (these moves are taken care
              of more eagerly above).
    \end{itemize}
    This completes the construction of $\pi_i$, and thus $\pi_y$ gives us a
    required accepting run of $\parampds$.
\end{proof}

\section{Complexity Lower Bounds}

\begin{theorem}
    The parameterised reachability problem for \NPDSs with a single global store
    is NP-hard, even when the stacks are removed.
\end{theorem}
\begin{proof}
    We reduce from SAT.  The encoding is as follows: $\master$ first guesses an
    assignment to the variables $x_1,\ldots,x_n$ (say).  He does this by writing
    $1_i$ or $0_i$ to the global store for each $1 \leq i \leq n$.  The $\slave$
    process has $n$ branches.  Along the $i$th branch it reads, and remembers in
    its control state, the value of $x_i$ written by $\master$.  Then, whenever
    a symbol $?_i$ can be read from the global store, $\slave$ reads it and
    writes $1_i$ or $0_i$ as appropriate.

    Then, also in its control state, $\master$ evaluates the boolean formula.
    When it needs to obtain the value of $x_i$. it writes $?_i$ to the global
    store and waits for a copy of $\slave$ to return the answer.  A unique
    control state is reached if the formula evaluates to true.  Hence, the
    defined parameterised reachability instance reaches this control state iff
    the formula can be satisfied.

    It is not immediately obvious how to evaluate the formula in the control
    state.  The technique is the same as in Hague and Lin~\cite{HL11}.  To
    evaluate a non-atomic formula, we store it as a tree in the control state.
    Evaluation uses a kind of tree automaton (the run of which is encoded into
    the state space).  The tree automaton navigates the tree in left most, depth
    first order.  First it moves down to the left most leaf.  This will be an
    atomic proposition.  The proposition is evaluated using the technique above
    and the value is passed up to the parent.  When first returning to a parent
    node, it is marked as seen.  If the node is a disjunction, and the value
    returned is $1$, then the automaton returns to the parent, also carrying the
    $1$, otherwise it moves down into the right subtree.  The automaton
    eventually returns from this tree with a value.  Since the node is marked,
    it detects that it has fully evaluated the disjunction and returns the value
    to the parent.  Evaluation is analogous for conjunction. Finally, a value is
    returned from the root.

    The evaluation above only introduces a polynomial number of control states.
    Because the tree is navigated in left most depth first order, there are a
    linear number of different markings (if the right hand subtree is not
    visited, we can simply mark all of the nodes in this subtree without
    affecting the execution).  Then, to keep track of the automaton, we attach
    the state of the automaton to the node of the tree it is at.  This is only
    polynomial since there is only one node marked by the automaton state at a
    time.
\end{proof}

\section{Proofs for Section~\refsec{sec:cfg2reg}}

\recaplma{lma:typed-cfg}{For all $w$, we have $w \in \langof{\typedg{\cfg}{x}}$
iff $x \in \spinetype{w}$.}
\begin{proof}
    First, assume $w \in \langof{\typedg{\cfg}{x}}$.  We show $x \in
    \spinetype{w}$.  Take the derivation tree of $w$ in $\typedg{\cfg}{x}$.  By
    definition, this tree has a path marked by a run of $\typenfa{x}$, such that
    $w$ is derived to the left (inclusive) of the path, and the empty word is
    derived to the right.  By replacing all non-terminals $A_q$ and
    $A_\varepsilon$ with their corresponding non-terminals in $\cfg$, and
    adjusting the applied production rules accordingly, we obtain a derivation
    tree of some word $wu$ containing a spine of type $x$.  Hence $x \in
    \spinetype{w}$, as required.

    In the other direction, consider the derivation tree $\dtree$ of $wu$ with a
    spine of type $x$ that witnesses $x \in \spinetype{w}$.  The spine induces
    an accepting run of $\typenfa{x}$.  Thus, we build a derivation tree of
    $\typedg{\cfg}{x}$ where all non-terminals and productions to the left of
    the spine are the same, all non-terminals and productions along the spine
    are annotated with the run of $\typenfa{x}$ and all non-terminals and
    productions to the right are replaced by their empty equivalent, e.g.
    $A_\varepsilon$.  This induces a derivation tree of $w$ in
    $\typedg{\cfg}{x}$ as required.
\end{proof}

\section{Lower Bounds on Automata Size}

We mentioned in the conclusion the problem of whether the doubly-exponential
size of the \NFA built from a very degenerate context-free language must be
doubly-exponential in the worst case.  We have been unable to obtain this lower
bound.  Since, in \refsec{sec:cfg2reg}, we construct a deterministic
finite-automaton, one may ask whether a result of Meyer and Fischer~\cite{MF71}
--- that there is a deterministic \PDS accepting a language whose corresponding
deterministic finite automaton is doubly-exponential --- can provide a lower
bound in the deterministic case.  Unfortunately, we provide a counter-example
below.  The language $I_n$ given by Meyer and Fischer is described as
follows\footnote{In the original definition, the word finishes with $\set{0,
1}^n$, though we believe this to be a mistake.  After this correction, the size
of the finite automaton is $2^{2^{n - 1}}$.  One could, of course, make other
corrections to preserve the $2^{2^n}$ claimed.}

\begin{quote}
    ``$I_n$ consists of words in $\set{0, 1, a_1, \ldots, a_n}^\ast \set{0,
    1}^{n-1}$ accepted by a deterministic pushdown store machine which operates
    as follows:
    \begin{enumerate}
        \item Copy the input onto the store until input $a_1$ is encountered.
              If $a_1$ does not occur, reject the input.

        \item Set $i = 2$.

        \item \label{alg:meyer-fischer:loop} If the next input is zero, pop the
              store until the first occurrence of $a_i$.  If the next input is a
              one, pop the store to the second occurrence of $a_i$.  If any
              other input is encountered, or the occurrences of $a_i$ are not
              found, reject the input.

        \item Increment $i$ by one.

        \item If $i \leq n$, repeat step~\ref{alg:meyer-fischer:loop}.

        \item If the digit on top of the store is $1$ and there are no more
              input symbols, accept the input.  Otherwise reject the input.''
    \end{enumerate}
\end{quote}

Intuitively, the input up to $a_1$ is interpreted as representing a binary tree
in post-fix notation (although the \PDS cannot enforce this with a small number
of states, hence even ``malformed'' trees are accepted).  After $a_1$, we see a
sequence of $0$s and $1$s tracing a path in the tree.  If this path ends on a
node labelled by a $1$, then we accept.  Since there are doubly-exponential
trees of depth $n$ labelled at the leaves by $0$ and $1$, we get that the
corresponding deterministic finite automaton must by doubly-exponential.

However, let $n = 3$ and consider the strong iterative pair 
\[
    \tuple{x, y, z, t, u} = \tuple{0 a_3 1 a_3 a_2,\ 0 a_3 0 a_3 a_2,\
    \varepsilon,\ 0 a_3 1 a_3 a_2,\ 0 a_3 0 a_3 a_2 a_1 1 0} \ .
\]
In the following, we underline the part of the input identified by the suffix $1
0$.  For $i = 0$ we have $xy^izt^iu = 0 a_3 \underline{1} a_3 a_2\ 0 a_3 0 a_3
a_2 a_1 1 0$ and for $i > 0$ we have 
\[ 
    xy^izt^iu = \ldots tu = \ldots 0 a_3 \underline{1} a_3 a_2\ 0 a_3 0 a_3 a_2
    a_1 1 0 \ .  
\]    
In both cases one can verify membership in $I_n$.

However, consider $i = 1$ and $j = 0$.  Then $xy^izt^ju = \ldots yu = \ldots 0
a_3 \underline{0} a_3 a_2\ 0 a_3 0 a_3 a_2 a_1 1 0$, which is not in $I_n$.
Essentially, the sub-tree given by $y$ violates the acceptance condition.  When
an occurrence of $y$ necessitated an occurrence of $t$, the automaton would
never read into $y$.  However, when $y$ and $t$ are disconnected, $y$ may not be
``protected'' by $t$.

    }{
        \input{shortrefs} 
    }

\end{document}